\documentstyle[aps,preprint,tighten,ifthen]{revtex}

\ifthenelse{\lengthtest{\topmargin<-50pt}}%
{\renewcommand\prl{1}}{\renewcommand\prl{0}}

\ifthenelse{\prl=1}{}{}
\ifthenelse{\prl=1}{\newcommand\sectionskip{}}%
{\newcommand\sectionskip{\smallskip}}
\ifthenelse{\prl=1}{\newcommand\paper{{Letter}}}%
{\newcommand\paper{Letter}}

\newcommand{\<}{\langle}
\renewcommand{\>}{\rangle}
\renewcommand{\d}{\partial}
\newcommand{\psibar}{\overline\psi}

\newcommand{\x}{{\bbox x}}
\newcommand{\p}{{\bbox p}}
\newcommand{\q}{{\bbox q}}
\newcommand{\dx}{dV}

\newcommand\sus{{\raisebox{0.15em}{$\chi$}}_{\raisebox{-0.1em}{\scriptsize $I5$}}}
\newcommand\muai{\mu_{\raisebox{-0.1em}{\scriptsize $I5$}}}
\newcommand\SU{{\rm SU}}
\renewcommand\O{{\rm O}}

\newcommand{\titleabstract}{%
\title{Pion Propagation near the QCD Chiral Phase Transition}
\author{D.~T.~Son$^{1,3}$ and M.~A.~Stephanov$^{2,3}$}
\address{$^1$Physics Department, Columbia University, New York, 
New York 10027}
\address{$^2$Department of Physics, University of Illinois, Chicago, 
Illinois 60607-7059}
\address{$^3$RIKEN-BNL Research Center, Brookhaven National Laboratory,
Upton, New York 11973}

\date{November 2001} 
\maketitle

\begin{abstract}
We point out that, in analogy with spin waves in antiferromagnets,
all parameters describing the real-time propagation
of soft pions at temperatures below the QCD chiral phase transition 
can be expressed in terms of static correlators.  This allows, in principle,
the determination of the soft pion dispersion relation on the lattice.
Using scaling and universality arguments, we determine the critical behavior
of the parameters of pion propagation. We predict that when
the critical temperature is approached from below, 
the pole mass of the pion drops despite the
growth of the pion screening mass. This fact is attributed to the
decrease of the pion velocity near the phase transition.
\end{abstract}%
}

\begin{document}

\ifthenelse{\prl=1}{\wideabs{\titleabstract}}{\titleabstract\vskip 2em}

{\em Introduction}.---%
Designed for the primary goal of discovering
quark-gluon plasma, experiments with collisions of large nuclei, such as
those pursued at the Relativistic Heavy-Ion Collider
(RHIC) at Brookhaven National Laboratory, also open up possibilities to
study strongly interacting matter at extremely high
temperatures and densities.  In particular, it is hoped that at least some
part of the rich phase diagram of QCD can be explored experimentally.
On the theoretical side, some problems, such as finding the equation of
state of QCD, can be effectively solved by using numerical Monte Carlo
techniques.  However, many important issues related to {\em real-time}
behavior and response of high-temperature 
strongly interacting matter cannot be
systematically studied by such methods.  This is because lattice
techniques rely on the formulation of quantum field theory in {\em
imaginary time}.  As a result, the question of what happens to the
hadron spectrum of QCD, which is important, for example,
for the understanding of certain
features of the dilepton spectrum observed in heavy-ion collisions,
cannot be easily answered in a reliable fashion.

Fortunately, many quantities characterizing the real-time behavior of
finite-temperature systems can be related, by exact identities, to
static (thermodynamic) functions.  The most familiar case is the
relation between the velocity of sound $u$, the pressure $p$, and the
energy density $\epsilon$: $u=(\d p/\d\epsilon)^{1/2}$.  A less
trivial example is that of spin waves in
antiferromagnets: it has long been known \cite{HHPR69} that 
at any temperature below the phase transition, 
at long enough wavelengths there
exist low-frequency spin waves which have a linear dispersion curve,
whose slope is given exactly in terms of static quantities.  

In this \paper,
we point out that, in thermal QCD, the dispersion relation of
soft pions can be determined entirely using static quantities.
Such quantities, in
principle, can be measured on the lattice.  Using this observation, we
show that the pion pole mass, which characterizes the propagation
of the collective pion modes, decreases as one approaches the critical
temperature, despite the well-known fact that the pion screening mass
increases in the same limit.

\sectionskip

{\em Pion dispersion from static quantities}.---%
From the point of view of symmetry properties, QCD at temperatures $T$
below or just above the temperature of the chiral phase transition
$T_c$ is similar to a Heisenberg antiferromagnet
\cite{PisarskiWilczek,RajagopalWilczek}.  With two light quarks ($u$
and $d$), QCD possesses an approximate chiral $\SU(2)_V\times
\SU(2)_A\simeq \O(4)$ symmetry, which is broken spontaneously to
$\SU(2)_V\simeq \O(3)$ by the chiral condensate.  This is similar to the
$\O(3)\to \O(2)$ symmetry breaking in antiferromagnets.
Moreover, the order parameter of QCD, the chiral condensate
$\<\psibar\psi\>$, is distinct from the conserved charges (the vector
and axial isospin charges), which makes the real-time behavior of
QCD similar
to that of antiferromagnets (but not of ferromagnets.)

By analogy with spin waves in antiferromagnets
\cite{HHPR69}, one can show that, at any $T$ below
$T_c$, the real part of the dispersion relation
of soft pions is given by
\begin{equation}
  \omega^2 = u^2 (\p^2 + m^2) \,,
  \label{omega}
\end{equation}
provided the quark masses are small enough.  We use the
following terminology: $u$ is the pion {\em
velocity} (although it is the velocity only when $m=0$), $m$ is
the pion {\em screening mass}, and the energy of a pion at $\p=0$,
$m_p=um$, is the pion {\em pole mass}.
At zero temperature, $u=1$, 
and the pole mass coincides with the screening mass.  At nonzero temperature,
there is no Lorentz invariance, and $u$ generally differs from 1
\cite{Shuryak:1990ie,PisarskiTytgat}.  Such pion modes with modified 
dispersion relation are termed ``quasipions'' in 
Ref.\ \cite{Shuryak:1990ie}.

We shall show that the parameters $u$ and $m$ can be determined
by measuring only static (zero-frequency) Euclidean correlators.
In particular, $m$ can be extracted from the long-distance behavior 
of the
correlation function of the operator 
$\pi^a \equiv i\psibar\gamma^5\tau^a\psi$,
\begin{equation}
  \int\!d\tau\,\dx\, e^{-i\q\cdot\x} 
{
\<\pi^a(x)\pi^b(0)\> 
\over \<\psibar\psi\>^2 }
=
   {
1
\over 
f^2} 
{\delta^{ab}\over \q^2+m^2} 
\,,
  \label{pipi1}
\end{equation}
where $x=(\tau,\x)$, $\psi$ is the quark field, $a,b=1,2,3$, $\tau^a$ are
isospin
Pauli matrices, ${\rm Tr}\,\tau^a\tau^b=2\delta^{ab}$, and
$\<\cdots\>$ denotes thermal averaging.  The integration over the
Euclidean time variable $\tau$ is taken in the interval $(0,1/T)$.
In Eq.\ (\ref{pipi1}) $\<\psibar\psi\>$ is the chiral condensate 
at zero quark masses.
Equation (\ref{pipi1}) also provides the {\em definition} of the 
temperature-dependent 
pion {\em decay constant} $f$.

The pion velocity $u$ is equal to the ratio of the above defined
pion decay constant $f$ and the axial isospin
susceptibility
$\sus$:
\begin{equation}
  u^2 = {f^2\over\sus} \, .
  \label{u}
\end{equation}
This is a close analog of the equation $c^2=\rho_s/\chi_m$ 
\cite{HHPR69} for the velocity of spin waves in  
antiferromagnets.
The axial isospin susceptibility $\sus$ can be defined as the second
derivative of the pressure with respect to
the axial isospin chemical potential [see Eq.\ (\ref{lquark}) below], 
or, equivalently, via the static
correlator of the axial isospin charge densities,
\begin{equation}
  \delta^{ab}\sus =
  \int\! d\tau\,\dx\, \< A_0^a(x) A_0^b(0) \>\,,
  \quad A_0^a \equiv \psibar\gamma^0\gamma^5{\tau^a\over2}\psi \,.
  \label{chi}
\end{equation}
The right hand side of Eq.\ (\ref{chi}) is free of short-distance
divergences in the limit of zero quark masses, when $A_0^a$ are densities
of conserved charges.

The derivation of Eqs.\ (\ref{omega})--(\ref{chi}) at non-zero temperature 
requires an analysis of the
hydrodynamic theory similar to the one performed in Ref.\ \cite{HHPR69}.
This approach will be presented elsewhere \cite{hot_pions}.
In this \paper, we use an intuitively simpler (but less rigorous) derivation
based on the effective Lagrangian approach. This
approach does not allow a correct treatment of dissipative effects, but will
be sufficient for our purpose. 
A somewhat similar approach has been used, for low temperatures, in
Ref.\ \cite{PisarskiTytgat}.


\sectionskip

{\em Derivation}.---%
%
Our strategy is to first write down the most general form of the effective
Lagrangian of pions and then relate its free parameters to the
correlation functions of QCD by matching the partition function 
${\cal Z}=e^{{\cal P}V/T}$ and
its derivatives in the effective and microscopic theories.
The quark part of the QCD Lagrangian at finite axial isospin
chemical potential $\muai$ is given by:
\begin{equation}
  {\cal L}_{\rm quark} = i\psibar \gamma^\mu D_\mu \psi 
  - (\psibar_L M \psi_R + {\rm h.c.})
  + \muai A^3_0 \,,
\label{lquark}
\end{equation}
where 
$M={\rm diag}(m_u,m_d)$ is the quark mass matrix.  The chemical
potential $\muai$ is coupled to the axial isospin charge $A^3_0$ defined in
Eq.\ (\ref{chi}).
For simplicity, we set $m_u=m_d=m_q$.

We assume that, in the infrared,
the pion thermal width is negligible compared to its energy.  This has been
seen in explicit calculations at low $T$ \cite{Goity}.  
The dynamics of the pions is
described, in this case, by some
effective Lagrangian ${\cal L}_{\rm eff}$, which we   
assume to be local, allowing expansion in powers of
momenta. This is equivalent to the assumption that the correlation
functions have only pole singularities, as in hydrodynamics.
To lowest order, the Lagrangian is fixed by symmetries up to
three coefficients, $f_t$, $f_s$, and $f_m$,
\begin{equation}
  {\cal L}_{\rm eff}= 
  {f_t^2\over4} {\rm Tr} \nabla_0 \Sigma\nabla_0 \Sigma^\dagger
  -
  {f_s^2\over4} {\rm Tr} \partial_i \Sigma\partial_i \Sigma^\dagger
  + {f_m^2\over2}
  {\rm Re }{\rm Tr} M\Sigma \,,
\label{leff:T}
\end{equation}
where $\Sigma$ is an $\SU(2)$ matrix whose phases
describe the pions. Because of the lack of Lorentz invariance, 
$f_t^2$ and $f_s^2$ are independent parameters.  

The chemical potential $\muai$ enters 
lowest-order effective Lagrangian (\ref{leff:T}) via
the covariant derivative $\nabla_0$ in a way
completely fixed by symmetries.  This can be seen by
promoting the $\SU(2)_A$ symmetry in (\ref{lquark})
to a local symmetry and treating
$\muai$ as the time component of the $\SU(2)_A$ vector potential
\cite{mu2}. The covariant derivative $\nabla_0$ is forced to have the form
\begin{equation}
  \nabla_0 \Sigma \equiv \partial_0 \Sigma - {i\over2}\muai
  (\tau_3\Sigma + \Sigma\tau_3)\,. 
\label{deriv}
\end{equation}

The structure of the Lagrangian (\ref{leff:T}) is analogous to that of
the
effective Lagrangian at finite (vector) isospin chemical potential
$\mu_{\raisebox{-0.1em}{\scriptsize $I$}}$
\cite{mui}.  A significant difference between the two cases is that
the QCD vacuum breaks the $\SU(2)_A$ (axial isospin)
symmetry {\em spontaneously}.
It is important to note, however, that the $\SU(2)_A$ {\em is} a
symmetry of the Lagrangian (at $m_q=0$), as good as the
$\SU(2)_V$.  The conservation of the axial isospin
current $A^a_\mu$ in the chiral limit makes the
consideration of finite $\muai$ entirely legitimate.

The pion dispersion relation following from Eq.\ (\ref{leff:T})
is given by Eq.\ (\ref{omega}) with
\begin{equation}\label{umf}
u^2={f_s^2\over f_t^2}  \quad\mbox{ and }\quad m^2 = {m_q f_m^2\over f_s^2}\,.
\end{equation}

Matching the second derivative of the pressure ${\cal P}$ with respect
to $\muai$ in QCD and in the effective theory, we find the relation
between $f_t$
and $\sus$:
\begin{equation}\label{chi_ft}
\sus = {\d^2 {\cal P}\over\d\mu_{I5}^2} = f_t^2\,.
\end{equation}
Together with the first of Eqs.\ (\ref{umf}) and $f=f_s$ (see below), 
this implies Eq.\ (\ref{u}).
The first derivative with respect to $m_q$ gives
\begin{equation}\label{pbpfm}
-\<\psibar\psi\> = {\d {\cal P}\over\d m_q} = f_m^2\,.
\end{equation}
Combining with the second of Eqs.\ (\ref{umf}), 
we derive the generalization of the
famous Gell-Mann-Oakes-Renner (GOR) relation to finite temperature:
\begin{equation}\label{GOR}
f_s^2 m^2 = -m_q\<\psibar\psi\>\,.
\end{equation}
Finally, we need to show that $f=f_s$.
We achieve this by treating $M$ as an external field, which we
parameterize as $M(x)=m_q\,e^{i\alpha^a(x)\tau^a}$.
Matching derivatives of $\ln {\cal Z}$, we find
\begin{eqnarray}\label{d2lnz}
&&\<\pi^a(x) \pi^b(0)\>
=
{\delta^2{\cal \ln Z}\over m_q^2\delta \alpha^a(x)\delta\alpha^b(0)} 
= f_m^4
\<\phi^a(x) \phi^b(0)\>\,;
\nonumber\\ 
&&\pi^a\equiv i\psibar\gamma_5\tau^a\psi,\quad
\phi^a(x)\equiv {\rm Re }{\rm Tr }\, i\tau^a \Sigma(x)/2\,.
\end{eqnarray}
Note that $\pi^a$ is defined in the microscopic theory (QCD), while 
$\phi^a$ is a field of the effective theory.

The correlation function of $\phi^a(x)$ can be calculated
by expanding the effective Lagrangian in (\ref{leff:T}) to second order in 
$\phi^a$.  We expect the result to match the correlator of $\pi^a$
only for small momenta, which means that we have to limit ourselves to
zero Matsubara frequency and small spatial momenta, e.g.,
smaller than the screening mass $m_\sigma$ of the order parameter
$\sigma=\psibar\psi$.
For the static correlator of $\pi^a$, 
by integrating (\ref{d2lnz}) over $\tau$ and using 
(\ref{pbpfm}),
\begin{eqnarray}
  && 
\int\!d\tau\,
{\<\pi^a(x)\pi^b(0))\> 
\over \<\psibar\psi\>^2
}
=
  \int\!d\tau\,\<\phi^a(x)\phi^b(0)\>
  \nonumber\\
  && =   {1\over f_s^2} 
  \!\int\! {d^3\q\over(2\pi)^3}\,
  { e^{i\q\cdot\x}\delta^{ab}\over \q^2+m^2} 
  =
  {1\over f_s^2}\, {e^{-m|\x|}\over4\pi  |\x|}\delta^{ab}\,.
  \label{pipi}
\end{eqnarray}
We see that, by measuring the large-distance ($|\x|\gg m_\sigma^{-1}$) static 
correlation function of the operator $\pi^a=i\psibar\gamma^5\tau^a\psi$,
we can extract two parameters of the effective Lagrangian (\ref{leff:T}):
the screening mass $m$, and $f_s$ which
coincides with the decay constant $f$ defined by (\ref{pipi1}).
The third parameter, $f_t^2$, coincides with the susceptibility 
$\sus$, which can also be 
expressed in terms of the static correlation function
in Eq.\ (\ref{chi}).  From Eq.\ (\ref{u}), we completely determine
the dispersion relation of soft pions.

\sectionskip

{\em Critical behavior}.---%
For the above results to be valid, pions must be the
lightest modes.  In particular, this requires $m\ll m_\sigma$.
If $m_q$ is very small, this
condition is satisfied everywhere below $T_c$, except for a region 
very close to $T_c$. As $T\to T_c$ from below and $m_\sigma\to0$, one can
ask the question: What is the critical behavior of the
parameters $u$ and $m$ when $T$ remains 
sufficiently far from $T_c$ so that the hierarchy
$m\ll m_\sigma\ll T$ is maintained?

Since $u$ and $m$ can be related to static
correlation functions, one should expect their critical behavior
to be governed by the same static critical exponents known from the
theory of critical phenomena.
We begin by considering the critical scaling of the
decay constant $f=f_s$. It is defined via the
behavior of a static correlator (\ref{pipi}) at distances larger than
$m_\sigma^{-1}$.  In the range of momenta
$m\ll |\q|\ll m_\sigma$
we have [see Eq.\ (\ref{pipi1})]
\begin{equation}\label{pb5p-corr}
  \int\!d\tau\,\dx\, e^{-i\q\cdot\x} 
  \<\pi^a(x)\,\pi^b(0)\> =
   \delta^{ab}{\<\psibar\psi\>^2\over f^2} 
{1\over \q^2} 
\,.
  \label{phiphi2}
\end{equation}
On the other hand, at distances short compared to the correlation
length, i.e., for momenta such that $m_\sigma\ll |\q|\ll T$, the
correlator of the order parameter $\psibar\psi$ has the following
scaling behavior:
\begin{equation}\label{pbp-corr}
  \int\!d\tau\,\dx\, e^{-i\q\cdot\x} 
\<\psibar\psi(x)\,\psibar\psi(0)\>
 \sim {1\over|\q|^{2-\eta}}\,.
\end{equation}
We also know that in this regime the correlators of $\sigma=\psibar\psi$
and $\pi^a=i\psibar\gamma^5\tau^a\psi$ are degenerate, since they are
related by the $\SU(2)_A$ symmetry, which is restored at $T_c$.
Thus the correlator (\ref{pbp-corr}) must match with the
correlator (\ref{pb5p-corr}) at the scale $|\q|\sim m_\sigma$.
This requires
\begin{equation}\label{f=mpbp}
  f^2 = A \ m_\sigma^{-\eta} \<\psibar\psi\>^2.
\end{equation}
The coefficient $A$ cannot be found from scaling arguments
but is finite and regular at
 $T_c$. The exponent $\eta$
in the
universality class of the $\O(4)$ sigma-model in $d=3$
dimensions (to which two-flavor QCD at $T_c$ belongs
 \cite{PisarskiWilczek})
is known: $\eta\approx 0.03$ (see, e.g., 
\cite{Baker-etal,RajagopalWilczek}). 

At $T\to T_c$ the scaling laws for the inverse correlation length
$m_\sigma$ and the order parameter
$\<\psibar\psi\>$ are also known from universality, 
\begin{eqnarray}
  \label{ms-scaling}
  m_\sigma & \sim & t^\nu ,\\
  \label{pbp-scaling}
  \<\psibar\psi\> & \sim & t^\beta ,
\end{eqnarray}
where $t=(T_c-T)/T_c$.
Thus we find
\begin{equation}\label{f-scaling}
  f^2\sim t^{2\beta-\nu\eta} = t^{(d-2)\nu},
\end{equation}
where in the last equation the relation
\begin{equation}
  2\beta=\nu(d-2+\eta)
\end{equation}
is used (recall that all scaling exponents can be expressed in terms
of two independent ones, e.g., $\eta$ and $\nu$).  From this point on,
we set $d=3$, so $f\sim t^{\nu/2}$.  The is the same as the Josephson 
scaling for the superfluid density in helium \cite{Josephson}.
Contrary to a naive expectation, $f$ scales
differently from the order parameter $\<\psibar\psi\>\sim
t^\beta$. The difference is numerically small, due
to the smallness of $\eta$.  In the $\O(4)$ universality class in
$d=3$, $\nu\approx 0.73$ and $\beta\approx0.38$ \cite{Baker-etal}.

Next, we point out that $\sus$ is finite at $T=T_c$, where it is degenerate
with the vector isospin susceptibility. The singular
behavior of $\sus$ is dominated by the mixing of $A_0^a$ with
operators linear or quadratic in $\sigma$ or $\pi^a$ in the
dimensionally reduced theory describing infrared modes $|\q|\ll T$.
Such a mixing, however, is forbidden by the $\O(4)$ chiral symmetry,
as well as by charge conjugation.
This is consistent with the lattice result that
the vector isospin susceptibility is finite at $T_c$ \cite{gottlieb}.  
Since $f_t^2=\sus$,
finiteness of $\sus$
invalidates a common assumption that $f_t\to0$ at $T_c$.
Note that above $T_c$ there are no propagating soft pion modes, so the
parameters of of the effective Lagrangian (such as $f_t$), and hence
Eq.\ (\ref{chi_ft}), lose their meaning, even as $\sus$ remains
well defined and finite.  This is not surprising if one recalls that,
as $T\to T_c$ from below, the domain of validity of the
Lagrangian (\ref{leff:T}) ($|\q|\ll m_\sigma$) shrinks away and
disappears at $T_c$.

Now we are ready to find the scaling of $u$.  Using Eq.\ (\ref{u}),
the scaling of $f$ in Eq.\ (\ref{f-scaling}), and the fact
 that $\sus$ is finite at $T_c$,
we find
\begin{equation}\label{u-scaling}
u^2\sim f^2\sim t^\nu.
\end{equation}
This means that the velocity of pions vanishes at $T_c$.

The scaling of the screening mass $m$ can be found from the 
GOR relation (\ref{GOR}) (recall that $m_q$ is assumed to be small):
\begin{equation}\label{mscr-scaling}
m^2 = -{m_q \< \psibar\psi \>\over f^2}\sim m_q\, t^{\beta-\nu}.
\end{equation}
In the $O(4)$ universality class $\beta<\nu$, which implies
that the static screening 
pion mass grows (at fixed $m_q\ne0$) as $T\to T_c$. 
This fact is already known from lattice simulations of QCD.

The {\em pole} mass of the pion, $m_p$, scales differently:
\begin{equation}\label{mp-scaling}
m_p^2\equiv u^2 m^2 = -{m_q \< \psibar\psi \>\over \sus}
\sim m_q\, t^\beta.
\end{equation}
This means the pole mass of the pion drops as $T\to T_c$.

For the formulas (\ref{u-scaling})--(\ref{mp-scaling})
to be valid it is necessary that 
$t\ll 1$.
However, for any $m_q\ne0$, these formulas
break down when $t$ is so small
that the condition $m\ll m_\sigma$ is violated.
Using Eqs.\ (\ref{ms-scaling}) and
(\ref{mscr-scaling}), we see that this happens when
$t\sim m_q^{1/\beta\delta}$ or smaller.
In the regime $t\ll m_q^{1/\beta\delta}$ 
the ``distance'' from the critical point
($T=T_c$, $m_q=0$) is controlled by $m_q$, but not $t$. The 
$m_q$ scaling of all quantities can be obtained starting from
\begin{equation}
\<\psibar\psi\>\sim m_q^{1/\delta}\quad \mbox{ at } t=0 \,.
\end{equation}
Comparing this to Eq.\
(\ref{pbp-scaling}), we see that $t$ and $m_q^{1/\beta\delta}$ have the
same scaling dimension. Using the scaling hypothesis we can easily
obtain the $m_q$ scaling by replacing $t$ with $m_q^{1/\beta\delta}$.
For example,
\begin{equation}\label{m-mq-scaling}
  m^2\sim m_q^{1-(\nu-\beta)/\beta\delta}, 
  \qquad 
  m_p^2\sim m_q^{1+1/\delta}.
\end{equation}
($m_p$ now has the meaning of the typical frequency of the pion mode
with zero momentum.  This mode may be overdamped in this regime.)
Both masses vanish as $m_q\to 0$ at $T=T_c$; however, for the
screening mass $m^2\gg m_q$, while for the pole mass $m_p^2\ll
m_q$. In particular, near the phase transition $m_p\ll m$.

The decrease of the pion pole mass may have interesting consequences for
heavy-ion collisions.  It is
the pole mass of a hadron, rather than its static screening mass, 
that affects the observed spectrum.
Within statistical models for hadron production, 
the drop in the pion pole mass would lead to an overpopulation
of pions 
at low momenta, provided the chemical freezeout
temperature $T_{\rm ch}$, at which the hadron abundances are fixed, 
is close to $T_c$.
For a crude estimate of this effect we use 
$\Lambda_{\rm QCD}\sim 200$ MeV
as the typical QCD scale in Eq.\ (\ref{m-mq-scaling}), 
 and $T_{\rm ch}\sim 170$ MeV as an estimate for the freezeout temperature
\cite{PBM}, which is indeed very close to $T_c$.
The shift of the pole mass near $T_c$ is approximately 
 $\Delta m\equiv m_p-m_\pi \approx
m_\pi ( (m_q/\Lambda_{\rm QCD})^{1/2\delta}-1)\approx -0.3 m_\pi$, where
$\delta\approx 5$, and the pion multiplicity at small momenta
is enhanced by roughly $\exp(-\Delta m/T_{\rm ch}) \approx 1.3$.  
This is a noticeable effect, although it is smaller than the
known contribution to pion overpopulation due to the feed-down from 
the decays of resonances \cite{PBM}.  This enhancement
is comparable to the effect of the
pion chemical potential $\mu_\pi\sim 50$ MeV induced by pion
kinetics after the chemical freezeout \cite{Shuryak}.

Another potential consequence of the fact that pion velocity decreases
 at $T_c$ is the possibility of Cherenkov radiation of pions by a hard
 probe moving through the hot medium created in a heavy-ion collision.

\sectionskip

The authors are indebted to D.~B\"odeker, N.~Christ,  
R.~Mawhinney, L.~McLerran, A.~Mueller, and
R.~Pisarski for discussions.  We thank RIKEN, Brookhaven National
Laboratory, and U.S.\ Department of Energy [DE-AC02-98CH10886] for
providing the facilities essential for the completion of this work.
The authors are supported, in part, by DOE OJI grants.  The work of
D.T.S. is supported, in part, by the Alfred P.\ Sloan Foundation.

\end{document}